\documentclass{article}
\usepackage[utf8]{inputenc}
\usepackage[english]{babel}
\usepackage{amsmath}
\usepackage{amsfonts}
\usepackage{graphicx}
\usepackage{hyperref}
\usepackage[numbers]{natbib}
\usepackage[nolist]{acronym}
\usepackage[affil-it]{authblk}

\author{Jonathan Spring
\thanks{This document is a preprint of a paper to appear in \emph{Computers \& Security}, \url{https://doi.org/10.1016/j.cose.2023.103191}}
}

\title{An Analysis of How Many Undiscovered Vulnerabilities Remain in Information Systems}

\begin{document}
\date{April 1, 2023}

\maketitle
\begin{abstract}
Vulnerability management strategy, from both organizational and public policy perspectives, hinges on an understanding of the supply of undiscovered vulnerabilities. 
If the number of undiscovered vulnerabilities is small enough, then a reasonable investment strategy would be to focus on finding and removing the remaining undiscovered vulnerabilities. 
If the number of undiscovered vulnerabilities is and will continue to be large, then a better investment strategy would be to focus on quick patch dissemination and engineering resilient systems.
This paper examines a paradigm, namely that the number of undiscovered vulnerabilities is manageably small, through the lens of mathematical concepts from the theory of computing. 
From this perspective, we find little support for the paradigm of limited undiscovered vulnerabilities. 
We then briefly support the notion that these theory-based conclusions are relevant to practical computers in use today.
We find no reason to believe undiscovered vulnerabilities are not essentially unlimited in practice and we examine the possible economic impacts should this be the case.
Based on our analysis, we recommend vulnerability management strategy adopts an approach favoring quick patch dissemination and engineering resilient systems, while continuing good software engineering practices to reduce (but never eliminate) vulnerabilities in information systems. 
\end{abstract}

\section{Introduction}
\label{sec:intro}

This paper will bring computing theory and security operations into conversation to answer the question ``How many undiscovered vulnerabilities are there in a piece of software?'' 
The answers to this question influence how both theoretical computer science and security operations should behave. 
The answer this paper's arguments support is that there are always more undiscovered vulnerabilities in any modern deployed software. 
I will also provide some possible directions for economic and research reactions to this situation. 

Eminent security writers such as Dan Geer \cite{geer2014_realpolitik} and Bruce Schneier \cite{schneier2014vuls} have weighed in on these questions. 
Both reach similar conclusions that are differently worded. 
Geer states, ``I believe that vulns are scarce enough for [cornering the global vulnerability market] to work.''  
Schneier states, ``But while vulnerabilities are plentiful, they’re not uniformly distributed\dots [P]ractices that eliminate many easy-to-find ones greatly improve software security.''
Dale Peterson and Josh Corman state that vulnerabilities are dense in medical devices, and mostly want to get on to discussing what to do about it \citep{peterson2018corman}.

The claim throughout all these assessments is, roughly, that good practices and smart policy can lead to software that is secure enough that it is fit for purpose. 
The statements in Peterson and Corman \cite{peterson2018corman} assess vulnerabilities to be dense because the devices are not created carefully. 
This implies the position that enough secure coding practice could be the end of vulnerabilites in software. 
Schneier and Geer take more pessimistic positions, but only slightly. 
All three positions leave the current software engineering and production paradigms intact and imply that the community can get to a place where software is secure.  
However, there is good reason to hypothesize that complete vulnerability listing will not be possible.
Namely, Rice's theorem~\citep{rice1953classes} states that all non-trivial semantic properties of programming languages are undecidable.

Other writers have also analyzed the question of software vulnerabilities. 
There is a discipline of measuring defect density in software engineering that was already well established in the late 1990s \citep{frakes1996software}.
This type of empirical analysis has been applied by security researchers since the mid-2000s (see for example \citep{ozment2006milk}). 
This paper will suggest that such metrics do not pose well-formed questions because---unlike the unresolved stances of Geer, Schneier, and Peterson---it appears to be clear from the following analysis that there are always more vulnerabilities to find (in Geer's terms, vulnerabilities are dense). 

This paper will relate vulnerabilities to the fundamental computer science understanding of the halting problem and Turing machines.
In so doing, we can understand what vulnerabilities being densely located in computer code actually means for security, software engineering, and program verification, and what it would actually take to eliminate vulnerabilities from software. 
Section~\ref{sec:theory} introduces the theoretical concepts of Turing machines and some light mathematical background.
Section~\ref{sec:security} introduces the information security concepts of vulnerability management, program verification, and fuzzing. 
Section~\ref{sec:vul-density} begins the contact between these two areas with a discussion of how vulnerabilities in software could meet the mathematical definition of density.
Section~\ref{sec:halt-vul} continues by discussing how program verification communities have interacted with the halting problem and what that means for pragmatic vulnerability discovery. 
Section~\ref{sec:practical} argues how the results from theory matter for practical, deployed software systems. 
Section~\ref{sec:cost} examines the ramifications of Sections \ref{sec:vul-density}~and~\ref{sec:halt-vul} on the cost of vulnerability discovery and defense.

\section{Primer on Theoretical Concepts}
\label{sec:theory}

Computing theory starts with a precise definition of computability. 
This definition is still useful for practical security managers because a computer can only calculate answers to questions that are computable. 
Computability is the study of what questions have computable answers. 
The practice of computer science has been to continually make computation more efficient.
This improvement has made calculating complex computable answers practical. 
However, no amount of practical improvement can make an uncomputable question answerable by a computer. 
Security managers should be familiar with these concepts to avoid asking uncomputable questions. 
Section~\ref{sub:turing-mach} introduces this definition of computability. 
Section~\ref{sub:halting} provides some mathematical background to give two important examples of questions that are uncomputable.
Section~\ref{sub:dense} completes the mathematical background by introducing the formal concept of ``density'' that some writers in information security use as a metaphor in discussions about vulnerabilities. 
Readers already familiar with these concepts may safely skip ahead to Section~\ref{sec:security}.

Section~\ref{sec:practical} argues that these formal models apply to modern software even if existing computers take up a finite amount of space and time on earth. 
In brief, any software with an interface such as a network connection or a human at a keyboard effectively has an infinite tape to read from and write to.
The network may have a bandwidth limitation, but this only caps rate, not size.
If software interfaces had composable security properties, a modern program could plausibly be treated as isolated, but this is not possible unless the interface is written in a language with much less expressiveness than Turing Machines (namely, deterministic context free languages) \citep{sassaman2013security}.
Section~\ref{sec:practical} also analyzes the case of a Turing Machine program running in isolation, with no interface to a network or human input, even though this restriction largely disqualifies all modern software.
The search space for vulnerabilities in even a modest isolated system is, to understate the situation, cryptographically secure.

\subsection{Turing Machines}
\label{sub:turing-mach}

The concept of a Turing machine was first introduced by Turing \citep{turing1936computable} as a tool in philosophy of mathematics to answer the question of whether there are mathematical objects that a person can define but nonetheless cannot be calculated. 
Turing \citep{turing1936computable} called his conceptual invention a ``computing machine,'' after the job title (``computer'') of a person paid to follow a set of instructions to calculate values for accounting, war planning, and other similar large, social projects. 
The cleverness of the construction is in its simplicity. 
The computing machine has two conceptual parts, which can interact in only one of four actions at a time. 
There is a strip of paper, the \emph{tape}, marked into squares that are either blank or not. 
The other part is the head, which can read, write with a pen, or erase.
The head takes up exactly one square on the tape and, after reading the current square, can do one of (1) move one square left; (2) move one square right; (3) erase the current square and make it blank; or (4) write a stroke in the current square to make it not blank.

The head part of the Turing machine is conceptually ``designed in such a way that at each stage of the computation it is in one of a finite number of internal \emph{states}'' \citep[p.~25]{boolos2002computability}. 
Each state is paired to an instruction. 
In this state, take this action and move to some other specific state.
For example, see \citep[p.~28]{boolos2002computability} for the specific description of states for a machine that doubles the number of strokes on the tape, written in 12 states.

Up to this point, making any of these into physical machines probably would not have been worth it. 
Turing's innovation here was to realize that he could construct a Universal Turing machine (UTM) where the initial state of the tape was a definition of states of a Turing machine. 
The central processing unit (CPU) in the device you are using to read this paper is a UTM. 
 When you loaded the software to read the paper, the CPU read in a set of instructions about a list of states and transitions that define the program. 
That is, it loaded the Turing machine represented by the program.
Once that was done, it started the program at the head of the tape with strokes and blanks (this file's representation in binary). 
The Turing machine then calculated what color each pixel on your screen should turn (represented by three numbers between 0 and 255 for the amount of red, green, and blue, respectively).
The values are between 0 and 255 because that is how many values eight squares on the Turing machine tape can represent. 
Since Universal Turing machines are extremely flexible, it is convenient to fix them into silicon and plastic as chips and reconfigure them millions of times during their useful life to perform different calculations. 

The problem for vulnerability management with this is the theorem that Turing was able to prove about Turing machines. 
This is known as the ``halting problem'' because it is about whether a given Turing machine $m$ with initial tape input $n$ reaches its final halted (stopped) state or not \citep[p.~38]{boolos2002computability}.
Section~\ref{sub:halting} will describe the background on this question. 
Section~\ref{sec:halt-vul} will explain why this presents a problem for vulnerability managers. 

\subsection{Halting Problem and Countable Infinity}
\label{sub:halting}

Several important questions in computing theory, such as the halting problem, depend on the concept of \emph{infinity}.
A simple understanding of infinity is that it describes something without bounds, or where there is always another; that is, something which is not finite. 
However, the tricky aspect of the concept of infinity is that it is not one concept. 
This section will differentiate two concepts: an infinitely large set that can be counted or enumerated and what it means for a set to be larger than that. 
This definition will be important because an uncountably large set is also uncomputable by a Turing machine.

The counting numbers are familiar to most school children. 
An \emph{enumerable} or \emph{countable} set is just one whose members can be enumerated.
That is, ``arranged in a single list with a first entry, a second entry, and so on, so that every member of the set appears sooner or later on the list'' \citep[p.~3]{boolos2002computability}. 
The set of positive integers $P$ is named by the list $1,2,3,4,5,\dots$, going on as long as you like. 

An \emph{enumerably infinite} set is any set whose members can be shown to be equivalent to, or a function of, the counting numbers $P$. 
While these lists in principle go on forever, for any specific $n^{\mathrm{th}}$ element we can list it. 
The tricky thing is that sets that intuitively seem smaller or larger than $P$ contain the same number of elements, since these sets all contain enumerably infinitely many elements. 

The following sets are all enumerably infinite (the official term is that they have the same \emph{cardinality}) \citep{boolos2002computability}:
\begin{itemize}
\item The positive even integers $2,4,6,8,10,\dots$
\item The whole set of integers $\dots,-2,-1,0,1,2,\dots$
\item The set of all ordered pairs of positive integers\\
$(1,1), (1,2),(2,1),(1,3),(2,2),(3,1),(1,4),(2,3),\dots$
\item The positive rational numbers $\frac{1}{1}, \frac{1}{2}, \frac{2}{1}, \frac{1}{3}, \frac{2}{2}, \frac{3}{1}, \frac{1}{4}, \frac{2}{3},\dots$
\item The set of ordered $n$-tuples for any integer $n$; this generalizes the set of all ordered pairs.
\end{itemize}

To show that a set is \emph{not} countable, we must demonstrate that there are always elements that are not in its listing. 
Computers work on the equivalent of one enumerable list, so anything uncountable a computer can at best approximate.

An example of an uncountably infinite list is the set of all sets of positive integers $P*$.
The set $P*$ contains the set of positive integers $P$; other infinite subsets, such as the even positive integers; and finite sets, such as $\{1,2,3\}$. We can convince ourselves this set $P*$ is not countable by showing a way to construct a genuine element of $P*$ that will never be in an enumerated list of the elements of $P*$. 
Since there will always be elements not in the list, the list is not enumerable or countable. 
The method for this is Cantor's \emph{diagonalization} algorithm \citep{boolos2002computability}.

Understanding whether each Turing machine will complete its computation and halt is equivalent to enumerating $P*$. 
The list of all possible Turing machines is countable. \citep[p.~36]{boolos2002computability}
Turing machines describe a method for computing properties or functions of the positive integers. 
But Turing machines are also positive integers. 
Conceptual constructions, such as numbers or logic, cannot prove properties about themselves \citep{sep-goedel-incompleteness}.
A demonstration of this for Turing machines is a construction of a diagonal function on the numeral representation of Turing machines \citep[p.~37]{boolos2002computability}.

Knowing whether a Turing machine will go through a finite number of state transitions is the same as knowing whether it will halt. 
The most efficient method for computing this fact is to run the Turing machine through its computations and see if it halts. 
To be able to calculate ahead of time whether it halts would mean defining a function $h(m,n)$ over the set of integers that represent Turing machines ($m$) and the input tape ($n$) such that $h(m,n)=1$ if $m$ eventually halts and 2 otherwise. 
If it were possible to construct a Turing machine $H$ to compute $h$, then we can construct a Turing machine $H'$ with a couple additions to $H$ that would \emph{both} halt \emph{and} not halt when given its own number as input \citep[p.~39]{boolos2002computability}. 
Therefore, the machine $H$ cannot exist; it is not possible to compute whether a Turing machine will halt in general better than by running that Turing machine and waiting to see.
That is, given a Turing machine, it is not possible to know a priori what the result of its computation will be. 

Turing machines are powerful conceptual machines, and the electronic computers designed to make them real can do an almost magical amount of work. 
We want to know whether listing all the vulnerabilities in a piece of software is equivalent to a countable or uncountable list.
Programs are Turing machines and vulnerable states are a subset of resulting states.  
Section~\ref{sec:halt-vul} will continue working with these conceptual tools and evaluate the extent they matter to practical vulnerability management.
There is good reason to hypothesize that complete vulnerability listing will not be possible.
This reason is known as Rice's theorem~\citep{rice1953classes}, which states that all non-trivial semantic properties of programming languages are undecidable (that is, uncountable). 
However, some claims about vulnerabilities are not whether they are countable or not, but whether they are \emph{dense} or not. 
This concept is related to infinite lists, but distinct, and Section~\ref{sub:dense} introduces it.

\subsection{Dense and Sparse Mathematical Spaces}
\label{sub:dense}

In computational logic, a set of ordered elements is \emph{dense} if there is always another element in between any two given elements.
Formally, a linear order is dense \citep[p.~152]{boolos2002computability} if it is a model of
$$ \forall x \forall y \left(x<y \rightarrow \exists z \left( x < z \wedge z < y \right) \right)  $$
and \emph{sparse} otherwise. 
The positive integers are sparse because there is no integer between 2 and 3. 
The rational numbers are dense because there is always some value in between. 
For example, $\frac{1}{2}$ between 0 and 1, $\frac{1}{3}$ between 0 and $\frac{1}{2}$, and so on. 
The positive integers are sparse and the rational numbers are dense even though both are countably infinite. 

\section{Primer on Security Concepts}
\label{sec:security}

The relevant security concepts for discussing this problem are the practices around management of vulnerabilities, especially discovery and response. 
Discovery usually refers to discovery of previously unknown vulnerabilities in a software product. 
Program verification and fuzzing are two categories of technology for discovering vulnerabilities in software. 
Response includes the detection of instances of known vulnerabilities as well as response actions to reduce risk. 

\subsection{Defining ``Vulnerability''}
\label{sub:vuldef}

What counts as a vulnerability is fairly broad. 
In practice, vulnerability managers may focus on vulnerabilities that are assigned a \ac{CVE-ID}. 
But the definition covers many more things besides \acp{CVE-ID}. 
Namely, a \emph{vulnerability} is
\begin{quote}
``a set of conditions or behaviors that allows the violation of an explicit or implicit security policy. Vulnerabilities can be caused by software defects, configuration or design decisions, unexpected interactions between systems, or environmental changes. Successful exploitation of a vulnerability has technical and risk impacts''
\citep{householder2020cvd}.
\end{quote}

The phrase ``explicit or implicit security policy'' is important. 
A security policy is ``a set of policy rules (or principles) that direct how a
      system (or an organization) provides security services to protect
      sensitive and critical system resources''~\cite{rfc4949}.
A crucial feature of security policies is that they are specific to systems and organizations. 
Security policies change over time as well.
Thus, whether a system contains a vulnerability depends on the context of which security policy applies.
This changeability is not simply because the system owner did not think of something. 
An \emph{implicit} security policy captures the scenario where a previously unthought side channel circumvents explicit security policy. 
Different organizations and systems have legitimately different security policies.
The difference is not merely on the surface due to lack of explicitness or lack of awareness of side channels.  
There are many cases with broad consensus on the security policy and, therefore, identification of vulnerabilities.
For example, if key management software has a flaw that discloses private/secret keys, the implicit security policy of the software design makes it clear this is a vulnerability. 
However, whether it is permissible for anyone on the Internet to query a DNS server depends on whether the software is deployed as an open recursive resolver (it is permissible) or a local resolver (not permissible). 

There are several different senses of the word vulnerability that are often clear to a security practitioner by context~\citep{spring2020managing}.
For example, a software product such as a word processor has a vulnerability.
Every installation of that word processor version contains an instance of the product vulnerability. 
The product vulnerability usually fits into a category of vulnerabilities, such as buffer overflows, as captured by \ac{CWE}. 
Discovery usually attends to product vulnerabilities whereas response usually attends to instances of vulnerabilities. 

The analysis in this paper focuses on implementation and configuration vulnerabilities in products and protocols.
Quite importantly, this leaves out design flaws.
Software may be implemented and configured without a flaw but still do the wrong thing, particularly if used in a context that was not considered during its design.
This point is important because it indicates that even after taking the most drastic compensatory steps identified in the following discussion, design flaws could still lead to situations in which software does the wrong thing at the socio-technical or human level, and that ``wrong thing'' may be a vulnerability.
As with all information security discussions, there is no panacea.

Vulnerability management ``includes services related to the discovery, analysis, and handling of new or reported security vulnerabilities in information systems'' as well as``services related to the detection of and response to known vulnerabilities in order to prevent them from being exploited'' \citep{csirtservices_v2}. 
Knowing how many undiscovered vulnerabilities remain in a piece of software would inform discovery, analysis, handling, detection, and response. 
Our assessment will focus on discovery and response. 
Section~\ref{sub:discovery} introduces the two most common discovery methods.
The remainder of this section introduces vulnerability response. 

\subsection{Vulnerability Discovery}
\label{sub:discovery}

Vulnerability discovery is about learning about vulnerabilities previously unknown to oneself. 
Many vulnerability managers accomplish this through mailing lists, Internet searchers, or other search methods. 
While these interpersonal methods are perhaps the most common, there would be nothing to discuss if analysis of programs and software could not discover new vulnerabilities. 
This section introduces the two most common methods of analysis to discover vulnerabilities in programs and software: program verification and fuzzing. 

Program verification is about assessing the properties of programs. 
If the properties a verification system is assessing are security properties, then program verification finds vulnerabilities by analyzing whether the program violates those properties. 
In software engineering practice, program verification techniques are best used as part of unit testing or other analysis at build time~\citep{ohearn2015categorical}. 
Program verification is easier if software and its correctness proofs are co-developed simultaneously, rather than trying to verify existing software \citep[\S8]{fisher2017hacms}.

Program verification was a mature field of study even in 1980~\citep{apt1981ten}. 
For at least as long, its practitioners have debated the nature of what program verification achieves~\citep{fetzer1988program}. 
One perspective is that programs are Turing machines about which one can conclusively prove properties based on appropriate logical methods, which is essentially treating programming as a branch of pure mathematics. 
An alternative perspective is that programs are an implementation on a physical machine, which one must conjecture properties to study via modeling and induction like a pond or a star system. 
An altogether separate view is that program verification may not be about proving program properties but rather checking for compliance with a specification.
On this specification view, logics are for clearly specifying and demonstrating properties of the specification of what the program is intended to achieve~\citep{lamport1983goodfor}.

Practical program verification contains aspects of each of these three perspectives. 
After 70 years, Turing's thesis and Rice's theorem continue to hold; there is no general-purpose method for proving properties of programs generally. 
What we have are myriad specific tools that are good at identifying specific properties in specific types of systems. 
One way to describe the situation is that program verification works best when the logical tools are designed and built to represent the relevant properties of the real-world system being verified~\citep{pym2018why}. 
So, for example, analysts have analyzed the \ac{TLS} protocol, with specific tools to check for specific security properties under specific assumptions~\citep{bhargavan2017verified}. 
To check whether some implementation of the \ac{TLS} version 1.3 protocol accurately and completely implements the protocol is a totally separate set of verification goals and tools. 
An analysis of whether some specific security properties are the best or fairest properties for the Internet community to adopt for an important protocol such as \ac{TLS} is mostly outside the scope of program verification.

Fuzzing, introduced in the 1990s, is a process of ``repeatedly running a program with generated inputs that may be syntactically or semantically malformed''~\citep[p.~1]{manes2019art}. 
Fuzzing is prominently used by researchers or adversaries after software is released because it is a kind of testing that only requires the evaluator to generate program inputs.
Due to its relative simplicity, it is used extensively by software developers before release as well. 

A good fuzzer can efficiently explore the program's input space to find inputs that cause unexpected program behavior. 
The fuzzing community has developed a variety of methods for accomplishing this, such as developing models of commonly used expected input types (for example, building a fuzzer for a PDF reader that only creates valid PDFs as input), predicting what groups or types of input will produce the same or similar output, or introducing some program verification techniques to build a model of the program's execution. 

Whether a systematic approach, such as program verification, or a random input mutation approach, such as fuzzing, is better or faster depends on the specifics of computation time~\citep{bohme2015probabilistic}. 
Computational efficiency aside, if we follow Pym et al.~\citep{pym2018why} and accept that specific program verification tools are suited to assessment of specific properties in specific systems, then there will always be a complementary place for fuzzing. 
Fuzzing makes fewer assumptions about what properties to search for and tries to find inputs that lead to unstable or undesired program states. 
For a defensive standpoint, fuzzing is particularly useful after program verification since we can search for unstable states outside the assumptions of the verification systems. 

One common problem for fuzzers is differentiating an input that causes a crash from an input that causes an \emph{exploitable} crash~\cite{manes2019art}.
A related challenge is when two exploitable crashing test cases should be called the same vulnerability or not, which often relates to whether one fix would eliminate both. 
But that definition pushes the problem on to defining whether changes to multiple source code lines or characters is one or two fixes. 
While this is of practical importance, for this paper we take definition assumptions most favorable to the paradigm of sparse vulnerabilities: We only consider as vulnerabilities those that are exploitable and consider all ``equivalent'' exploitable crashing states as one vulnerability. 

A practical problem for analyzing fuzzing results is generalization. 
Suppose we run a fuzzer against a program for 300 hours, and it discovers no vulnerabilities.
It is unclear what we should believe about how many vulnerabilities remain undiscovered in the program. 
In ecology, a researcher can answer similar questions by capturing two samples of fish in a lake and measuring how many fish were caught both times~\citep{pollock1991review}. 
But such statistical measures do not work for vulnerabilities from fuzzing campaigns because (1) it is not obvious when two crashing test cases are in fact the same vulnerability, and (2) the statistics require an assumption of statistical independence between the two capture events that program outputs do not meet. 

Fuzzing, like program verification, is a useful practical tool for vulnerability discovery. 
However, also like program verification, it provides little to no empirical evidence about what vulnerabilities may yet continue to be undiscovered in a program. 
Section~\ref{sec:halt-vul} returns to the question of undiscovered vulnerabilities after Section~\ref{sub:response} discusses actions in response to known vulnerabilities. Section~\ref{sec:vul-density} clarifies the definition of a vulnerability relative to the necessary mathematical concepts. 

\subsection{Vulnerability Response}
\label{sub:response}

Vulnerability response is a part of security operations.
Security operations are the practical administration of information systems that support a security architecture and delivery or recovery of security services \citep{rfc4949}.
Specifically, vulnerability response includes detection of deployed vulnerable systems and actions to mitigate or remediate risk due to detected systems~\citep[\S7.6]{csirtservices_v2}. 

Vulnerability detection occurs in at least three ways: scanning, penetration testing, and \ac{IDS} logs. 
Scanning, as usually done, requires a database of software versions and what vulnerabilities affect each version.
The scan is usually a software fingerprinting and asset management effort followed by a database lookup of which vulnerabilities affect each software version. 
This kind of effort is dependent on a good database of vulnerabilities. 
The \ac{NVD} is often used in practice, which contains all the \acp{CVE-ID}. 
Scanning, therefore, often focuses on vulnerabilities which have been issued a \ac{CVE-ID}.   

Penetration testing is when ``evaluators attempt to circumvent the security features of a system,'' often within some constraints or fixed starting conditions~\citep{rfc4949}.
Some penetration testing tools, such as Metasploit, function as a vulnerability scanner (as above) connected to exploit modules for the associated vulnerability.
The evaluator often works from a different set of vulnerabilities than the \ac{NVD} and tries to pivot to different scanning points to test systems that might not appear in a fixed external scan.
The evaluator may also search for common weaknesses in servers. 
Overall, a vulnerable system is any system on which the evaluator could circumvent security features. 

\ac{IDS} logs may detect adversary scans or indicators of successful attacks. 
Adversary scans may detect vulnerable systems, and the responses that indicate system vulnerability should also be captured in the logs. 
Indicators of compromised systems are not specific vulnerability detectors, but they indicate systems that likely have some vulnerability in them since they are compromised. 
There is a perhaps definition-based debate as to whether a weak or reused password that allows an adversary access is a ``vulnerability.''
But regardless of how that definition is split, a compromised system manifests some condition or behavior which calls for a response action. 

Actions to reduce risk from vulnerabilities fall into two categories: mitigation and remediation~\citep{dodi_8531_2020}. 
\emph{Mitigation} is reducing the impact of a vulnerability without eliminating it. 
Examples include restricting network access to vulnerable components or limiting access to a vulnerable component through software configuration changes. 
\emph{Remediation} is an action that eliminates or removes the vulnerability. 
Examples include applying a patch or decommissioning the vulnerable system.

\section{Definition of Dense or Sparse Vulnerabilities}
\label{sec:vul-density}

Before we can analyze whether vulnerabilities are dense, we have to make this metaphorical connection to number theory a bit more explicit. 
Security practitioners define a vulnerable state as a set of conditions or behaviors.
In the language of Turing machines, a single vulnerability (say, as identified by \iac{CVE-ID}) is one or more states of the Turing machine that lead to those conditions or behaviors. 
There are various formal attempts to define vulnerability in the context of software. 
Dullien~\citep{dullien2017weird} treats a specific software program as an emulator for an intended ``implicitly specified finite state machine'' and treats exploits as a discrepancy between what the software does and this intended machine.
Primiero et al.~\citep{primiero2019mal} adapts the concepts of malfunction and dysfunction from the broader technology studies literature.
But as with our definition of vulnerability introduced in Section~\ref{sub:vuldef}, these definitions hinge on some discrepancy between the actual state of the computer and the intended state of the computer. 
The intention of this section is to find an adequate formalism to discuss the claim that ``vulnerabilities are dense.'' 
The section concludes that the metaphor breaks down too much to be formally useful.

Vulnerabilities per se do not constitute an ordered set, and so the term \emph{dense} does not properly apply. 
But we can take the colloquial meaning of ``there is always another element'' and define what that might mean for vulnerable states of a Turing machine. 
For a conceptual Universal Turing machine, Section~\ref{sec:halt-vul} will argue that enumerating all vulnerabilities is equivalent to the halting problem.

For vulnerabilities being sparse or dense to be a meaningful question, we need a measure by which states generally are dense and vulnerable states might not be. 
We can model each state $s_n$ of the Turing machine as a rational number $n$ such that $0 \leq n \leq 1$, where $n$ models how well the state of the program adheres to the relevant security policy.
This model creates a preference order over the states of the program, where $1$ is most preferred because it is in alignment with the security policy and $0$ is least preferred because it is completely unaligned with the security policy. 

However, this mapping to a subset of the rational numbers is not particularly interesting.
Since any subset of a dense linear order is also dense, if there are any vulnerabilities at all, then there are dense vulnerabilities. 
That is, either every state is completely in line with the security policy ($n=1$), or there is always another state with a smaller preference ranking (more vulnerable). 

One vulnerability identified by a CVE-ID is usually not one single state of the program.
A CVE-ID usually identifies a set of states that are functionally equivalent in violating a security policy. 
Therefore, the question is whether the set of such sets is dense, not the set of states themselves.  

Therefore, we should prefer a measure over sets of states, rather than of individual states. 
We can simplify the values for $s_n$ to be a binary $f(s_n)\in\lbrace0,1\rbrace$ that represents whether the state satisfies the security policy ($1$) or not ($0$).  
A listing of all the states with value $1$ is essentially the definition of the security policy. 
Although Schneider~\cite{schneider2000enforceable} examines different concerns, the formalism used here is largely compatible with the formalism used to analyze  enforceable security policies in \citep{schneider2000enforceable}.
If we had that complete list, then any actual set of states of the program could receive a score based on the proportion of states that meet the policy versus the total states in question. 
There are several aspects of this modeling choice that would be difficult in practice, such as creating the infinite list of states and assigning each a value. 
The ramifications of this difficulty are discussed further in Section~\ref{sec:halt-vul}.
However, for now, let us say that we can score a set of states ($S$ with $s_n\in S$) as a rational number $a$ where $a=\frac{|s_n\in S : f(s_n)=1|}{|S|}$, with $|S|$ denoting the cardinality (size) of a set.

This definition puts the security score $a$ of a set of states on the rational interval $\left[0,1\right]$.
However, it's not clear how this is useful in modeling sets of states of the program. 
If there is a provably finite set of states where $a=1$, which is something that program verification tools can regularly produce, then it does not inform about the question of whether there are other sets of states with a lower value of $a$ (that is, a set of states where some are insecure). 

Alternatively, we could try to reason about the set of all states of the program. 
However, unless we can prove that the program will halt, we must assume an infinite number of states. 
If $|S|=\infty$, then $a=0$ due to the convention that any integer divided by $\infty$ is $0$. 

If we take subsets of the sets of program states $S_0,S_1,\dots,S_i,\dots$, then we end up analyzing all possible combinations of sets of states, which is the power set of $S$, notated $\mathcal{P}(S)$.
The values of $a$ over each set $S \in \mathcal{P}(S)$ will be dense if there are infinitely many vulnerable states. 
Therefore, the answer to whether this measure is dense depends on the question of whether vulnerabilities are dense, so it is not useful in answering our question of interest. 
If vulnerabilities were not dense, then there would be a maximum non-1 value of $a$ in $\mathcal{P}(S)$ at $\frac{v}{v+1}$ where $v$ is the number of vulnerable states, as long as there is at least one secure state.

Given these modeling results, perhaps it does not make sense to try to formally equate the metaphor of dense vulnerabilities to a mathematical concept of a dense linear order. 
If vulnerable states do not meet the definition of \emph{dense} in Section~\ref{sub:dense}, there still may ``always be another one.'' 
After all, there are countably infinite positive integers $P$, even though the positive integers are not dense. 
Unfortunately, this means we cannot rely on structured reasoning rules that an ordered space would supply.

\section{Halting and Vulnerabilities}
\label{sec:halt-vul}

Given these problems with numerical representation of vulnerable states, the problem should be modeled more generally. 
This section will argue that a solution that is appropriately general to avoid the problems described in Section~\ref{sec:vul-density} will instead be subject to the halting problem. 

Let's take a model of states of the Turing machine where the states are labeled by a natural number $i$, but there is no sequence or other meaning to $i$, it is just a label. 
Let us hypothesize that there is a function $v$ where $v(m,n,i)=1$ if $i$ is a secure state in Turing machine $m$ started with input $n$, otherwise $v(m,n,i)=2$. 
This modeling choice provides space to ask the relevant question: For any given Turing machine $m$ with input $n$, is the number of vulnerable states finite or infinite?

If vulnerabilities represent sets of states, then the set of these sets is uncountably infinite if both (A) the size of the sets of states is unbounded and (B) the number of sets of states is unbounded.
If one of these two conditions holds, then vulnerabilities are countably infinite. 
From this conceptual lens, countably infinite might be an acceptable security outcome because it should be possible to create a program that lists all the vulnerable states.
Uncountably many vulnerabilities would indicate there are always more and the best one could do, a priori, is to estimate some vulnerabilities in a Turing machine. 
This situation would mean that formal verification of a Turing machine could never, in principle, discover all the vulnerable states. 

The output of the function $v$ would be a series of $1$s and $2$s, for example: 
$$1,1,1,2,2,1,1,1,1,2,2,2,2,\dots$$
The first challenge with analyzing $v$ would be enumerating the input. 
To answer part (A), the question amounts to asking whether, given a $2$, if the next value will be a $2$. 
But to know whether $i$ is a state of $m$, one must enumerate all the states of $m$, which intuitively appears to be the halting problem again. 
Since all Turing machines are bound by the halting problem, for the general case we have to treat the sets of states of the machine as unbounded.
Section~\ref{sec:practical} will constrain this analysis, since computers that humans can build so far have bounded (rather than limitless) memory. 

This function $v(m,n,i)$ takes similar inputs as the proposed halting function $h(m,n)$.
We can make an argument from contradiction that $v$ does not exist. 
Assume the function $v$ exists.
For $v$ to be well defined, we must know the range of $i$ for which there are states of the machine $m$ for input $n$. 
Either there is a finite integer value $K$ for the largest $i$, or $i\in \mathbb{N}$ and $i$ is countably infinite. 
If $K$ exists, it is the halting state of $m$ with input $n$. 
If $i \in \mathbb{N}$, then $m$ does not halt with input $n$. 
That is, if $K$ is finite, then the function $h(m,n)=1$, and if $i \in \mathbb{N}$, then $h(m,n)=2$. 
As described in Section~\ref{sub:halting}, $h$ is not computable. 
But if we could compute $v$, then we could compute $h$. 
Therefore, $v$ must not be computable. 

Let us make a less general attempt.
Rather than one function that determines whether any Turing machine is vulnerable, what if there are different functions for some specific machine $m$ with specific input $n$.
That is, $v_m^n(i)$ might be possible for some $m$ and not others. 
In practice, program verification works better when it is specific to a type of system and type of vulnerability \citep{pym2018why}.
Requiring a different function for each machine $m$ is not exactly the same as focusing on a type of system, but it is driving at the same idea of specialization. 

Let us imagine a machine $m$ where it happens to halt, and we know this because we have run it and observed it to halt with input $n$ at state $K$. 
Then it would be possible construct the function $v_m^n(i)$ for $i \in K$. 
``Possible'' means there is no mathematical definitional contradiction.
When we declare the definition of $v_m^n(i)$, we are essentially declaring a security policy for the Turing machine.
At this point, there is another important feature of the function to consider: uniqueness. 

The information about the input state $i$ is useful if there is exactly one output, $1$ or $2$, vulnerable or not vulnerable.
If $v_m^n(i)$ is a function, as we've defined, and $v_m^n(i)$ is unique, then this is true. 
However, vulnerabilities are relative to a given security policy (see Section~\ref{sec:security}).
Therefore, if a different principal asserts a different security policy for machine $m$ with input $n$, we might have ${v_m^n}'(i)$ where for some values of $i$ it is the case that $v_m^n(i) \neq {v_m^n}'(i)$. 
This multiplicity is not a problem when declaring function definitions (that is, security policies), but it is a problem with algorithms that search for a valid function definition as if there is just one unique valid function to find when in fact there can be more than one.

This section has demonstrated that in the general sense of Turing machines, there is not a general method to discover (compute) vulnerabilities.
Specialized mathematical functions for specific Turing machines with specific inputs amount to declaring a security policy. 
Therefore, an important question is how pragmatic it might be to find a ``good enough'' approach between these two extremes, which will be analyzed in Section~\ref{sec:cost}.
First, Section~\ref{sec:practical} argues that this section's conclusions from the general modeling choice of Turing machines is useful and applicable to real modern systems.   

\section{Applicability to Modern Systems}
\label{sec:practical}

Modern systems are finite, unlike Turing machines. 
This section argues that modern systems are complex enough that exhaustive analysis is hopelessly out of reach; therefore the infinite model of a Turing machine is, for our intents and purposes, an adequate model of modern systems.
This argument will be sketched through an estimate of how to exhaustively check for vulnerabilities in a modern computer system. 
 
A vulnerability is a set of conditions that lead to a security policy violation. 
One thing many of these sets of conditions have in common is mismanagement of some system resource. 
The resource may be allocations of physical memory, filesystem pointers, memory pointer variables, network bandwidth, etc. 
Although there is a diverse set of resources that a computer system manages, one might question why resource management and allocation is so hard. 

One answer to this question is that the system cannot know a priori when a program (that is, a Turing machine) will stop needing resources, because the system cannot know when it will stop. 
Computer engineers have created a beautiful, clever, and diverse web of heuristics to estimate how many resources a program should be allocated. 
Shared physical memory is one such. 
While 250 programs may be running at any given time in memory, it is unlikely that any given one of them will need $\frac{1}{250}$ of available memory. 
Memory could, hypothetically, be allocated statically based on a maximum number of programs, such as 250, with firm boundaries where programs could not reference memory locations outside their allotment. 
But it has proven much more useful to dynamically allocate memory in a general-purpose way because, normally, a program can be trusted to just ask for more memory if it needs it.
This usually works. 
But the trust sometimes fails---a program is allowed to read memory that was not rightfully allocated to it, and we get vulnerabilities like ``Heartbleed'' \cite{heartbleed2014}. 

So as long as our computers are subject to the halting problem, we should expect to find more resource mismanagement vulnerabilities. 
One natural question, then, is to what extent are our current computers actually subject to the halting problem? 
The Turing machine model requires an infinite tape to read from and write to. 
Our computers are not infinite, so it seems like the halting problem should not apply to the software Turing machines running on it. 

An alternative modeling choice, such as pursued in \citep{dullien2017weird}, is to model a modern system as an implicitly specified finite state machine. 
The term ``implicitly specified'' does a lot of work here, the idea being the program written by the developer implicitly specifies a finite state machine, since it is written for a real computer to run, even though the programming language it is written in is a Turing Complete language. 
This allows \citep{dullien2017weird} to define exploits as a mismatch between this intended, implicit specification and the actual program execution. 

However, the Turing Machine model fits modern software better for our purposes. 
Any software with an interface such as a network connection or a human at a  keyboard effectively has an infinite tape, the software is not limited to local \ac{RAM}.
The network may have a bandwidth limitation, but this only limits how fast the tape can be used, not how large it is. 
So while we could get into discussions about how the program is likely to stop by the time the sun turns into a red giant and consumes the earth, that is not constructive. 
Since a network connection to the modern Internet allows for indefinite reads and writes interacting with arbitrary other processes (including physical and human processes), the best model for any interfaced program is an infinite-tape Turing machine.

We can consider a program running in isolation, with no interface to a network or human input; note this restriction would largely disqualify all modern software.
Isolated programs could be modeled as finite state machines with an unmanageably large set of possible states and transitions.
This is the case even if we limit the analysis to physical \ac{RAM}, and disallow mapping of virtual memory on disk, and disallow reads and writes from memory to disk in general. 
With a modest 4~GB of \ac{RAM}, available on any mid-tier 2020 smart phone, most of the memory is available for any program to use. 
If we say a program might be allowed to use 3.5 GB, there are $\left(2^8\right)^{3.5\cdot10^9} = 256^{3,500,000,000}$ possible states.  
Since $log_{10}256=2.408$, this is equivalent to $10^{8,428,839,878}$ states.  
32 bytes ($2^{256}\approx10^{77}$)  is a cryptographically secure search space \cite[p.~59]{nist800-57r5}.
So unless a program's memory footprint can be reduced to less than the equivalent of 14 bytes, it will not be practical to check through all of them ahead of time to ensure all the configurations are allowed by the security policy. 
As long as the program is written in a Turing Complete language, one cannot rule out any of these possible states a priori. 

Reducing program states to larger equivalence classes in order to do some logical verification is one interpretation of the practice of program verification.  
For example, a common tactic in program verification is to express a \texttt{while} loop as a logical sentence with a loop invariant where that sentence formally entails that the loop will terminate. 
This tactic creates an equivalence class among all the memory states the \texttt{while} loop can create and captures them in the rules of entailment and the loop invariant. 
With a 1 kB numerical array in a loop, the $2^{8192}$ possible memory states could be reduced to one statement expressible in a couple dozen bytes (characters).
This kind of reduction in state to search is quite powerful.

Program verification is quite useful at finding specific kinds of flaws, especially when the verification system is tailored specifically to one narrow kind of flaw \citep{pym2018why}.
Therefore, program verification works well in specific contexts, and has been demonstrated to be useful for preventing several specific classes of program flaw. 
In support of this conclusion, an analysis of 1100 Android applications found 22\% of non-trivial methods had at least one decidable property~\citep{barr2019sub}.
However, as long as those methods (which may be considered to be less expressive than Turing Machines) interface with Turing Complete methods and systems in an unrestricted way, as almost any real modern system does, the more expressive Turing Machine model is more appropriate.

Verification cannot solve the whole implementation problem, even with myriad different specialized verification modules, as long as Turing Machines are the best model. 
This is besides the fact that vulnerabilities may be due to design errors. 
There is some work in verification of the properties of protocols, such as TLS. 
Microsoft has successfully deployed formal parser verification in network stack handling for the Windows Virtual Switch~\citep{swamy2022hardening}, for example, which eliminates many classes of implementation vulnerabilities in that component. 
This does not comment on whether TCP as specified has flaws or ambiguities. 

A world where most software architecture plans are verified would help and we should move towards that.
However, a verified implementation of a bad design will have vulnerabilities that are outside the scope of what the verification can find. 
There is some evidence that file formats diverge into distinct but related file formats as different usage and parsers are developed \citep{cowger2020icarus}. 
The authors demonstrate the feasibility of format discovery supporting formal parser enforcement, but one finding is that the authors discover vulnerabilities in the formats in use. 
Analysis of format specifications in diverse fields such as industrial control system network protocols, PDF, \ac{ELF}, and data description languages have found vulnerabilities or ambiguities in the specification~
\citep{anantharaman2022protecting,mundkur2020parsley}.
The scope of program verification is generally limited to properties of the program, not the specification. 
Program verification is in practical use to make software better, but it is not within its scope to eliminate all vulnerabilities.
 
In our hypothetical 4 GB of memory, we might think of a successful program verification system as reducing all the possible states of, say, 1 kB of memory to a single verifiable statement.
This reduction would be quite successful if we can find one statement for each 1 kb of memory. 
Maybe the 1 kB represents a loop over the integers, and we can reduce that loop to a Hoare triple representing inputs and outputs of the code segment. 
What the correct statement is requires some computation, so it is not free to find, but once we have it there is a great complexity reduction. 
In the case where each memory chunk is treated as independent, it is as if the problem has been reduced from searching  $10^{8,428,839,878}$ states to searching the $10^{1,028,912}$ possible interactions between statements.
Both of these numbers are so much larger than $10^{77}$ that the distinction makes no difference; for both we must assume there remain undiscovered implementation vulnerabilities in the program. 
If the logic statements are not independent, then it is easier to get a conclusive result from the statements but harder to find a set of statements that are both valid and accurately models what the program actually does. 

Some program verification techniques are \emph{composable}, which means changes to one part of memory demonstrably do not affect other regions of memory. 
This property helps dramatically with scaling the computation of proofs by parallelizing computation and reanalysis of a slightly modified program. 
Such systems can make pragmatic claims about specific memory safety properties of programs \citep{calcagno2007local}.
Composable logics for program verification are clearly a good engineering choice for practical program verification, but they need to be specialized enough that we arrive back at the problem of what properties should be verified. 
In order to reduce the complexity of the problem, logic tools---based on, for example, temporal logic \cite{lamport2002specifying} or separation logic \cite{calcagno2007local}---do not search for security flaws in general. 
They also do a complex amount of syntactic and semantic manipulation such that the validation failures are often not actually flaws in the program, even in good systems \cite{calcagno2015moving}.
Some of this manipulation amounts to estimating an expression of the program in a sub-Turing language, such as finite state machines. 
But as long as the program is not actually written and specified in an appropriate sub-Turing language, the program verification will make mistakes at this phase.  
While it is pragmatic and useful to be confident an analysis has identified any potential null pointer dereferences, this accomplishment is far from analyzing all possible program states for all possible vulnerabilities.  

Sassaman et al.~\cite{sassaman2013security} argue that input validation in particular is not decidable for any program written in a Turing-complete language. 
A related result is that for security properties to be composed when system components are composed, those components must be written in a sub-Turing Complete language. 
A consequence of this result is that, while fuzzing and program verification are useful, any (Turing Machine) program that interacts with other programs will have vulnerabilities at the interface. 
Pragmatically, every modern computer has many such interactions. 
Just to even get started, the firmware has to interface with the bootloader which has to interface with the operating system. 
With regard specifically to input validation, ``well-specified input languages [can be made to be] DECIDABLE'' \citep[p~5][emphasis original]{sassaman2013security}, but many in use today (such as PDF and HTML) are undecidable languages. 
One example of an engineering decision that could eliminate input-validation vulnerabilities is to write the interfaces to use a sub-Turing language, specifically a deterministic context-free language. 
Any system written in a more expressive language than this, such as a Turing Machine, has input validation errors that are currently undiscovered. 
Any grammar where a length field influences parsing behavior (which includes basically every modern file system and network protocol) is at least context-sensitive \citep[p~7]{sassaman2013security}, and so should be expected to have undiscovered input-parsing vulnerabilities. 
DARPA's \href{https://www.darpa.mil/program/safe-documents}{SafeDocs} program appears to be investigating a similar approach to input validation assurance. 
 
Although vulnerabilities are not technically infinite in a 4GB system without an interface, it is not practically feasible to exhaustively check every possible state of a moderately complicated program for whether it violates every security policy. 
Almost every modern system breaks these modest assumptions anyway, as having an interface to a network or a human provides an effectively infinite tape and makes a Turing Machine the appropriate model. 
If the interface had composable security properties, a modern program could plausibly be treated as isolated, but this is not possible unless the interface is written in a deterministic context-free language (such languages are even less expressive than finite state machines) \citep{sassaman2013security}.
Program verification can usefully check many program states about some specific security policy \citep{pym2018why}.
And multiple program verification techniques can be used to check about different specific security policies. 
But the search space of possible states is too large to exhaustively check a Turing-Complete program, even with the help of fuzzing and program verification. 
Therefore, everyone should expect the result that Turing Machines always have more undiscovered vulnerabilities to apply to any modern system. 

There is a practical question that remains open: 
Can the effort an attacker must spend to find another vulnerable state be increased to the point that it costs the attacker more time or money than the knowledge of the vulnerable state is worth? 

\section{Cost of Vulnerability Discovery}
\label{sec:cost}

As introduced in Section~\ref{sub:discovery}, the effort to find previously unknown vulnerable states of a program is \emph{vulnerability discovery} \citep{csirtservices_v2}. 
Both attackers and defenders conduct vulnerability discovery.
Defenders use the information to make the vulnerable states unreachable, while attackers use it to force the program into vulnerable states. 
The overall goal of a secure software engineering practice would be to find and fix every vulnerable state before an attacker does. 
If vulnerabilities are dense or never-ending, then the software engineer cannot close all the vulnerabilities before finishing the software. 
If vulnerable states are so numerous as to be effectively infinite, then the result is the same. 
As Section~\ref{sec:halt-vul} argued, there is good reason to believe vulnerabilities are effectively unlimited.

In a world where software engineers cannot close all vulnerabilities in a program, how much effort should they put into closing some of them? 
This question blossoms into a complex set of interdependencies. 
The development team should expect some vulnerabilities to be found by external parties; therefore, they should have a vulnerability management function that can receive reports, triage them, and publish updates \citep{csirtservices_v2}.
Prioritization during triage is a complex topic in its own right \citep{spring2020ssvc}.
And the market for software expresses some economic inefficiencies that often push suppliers of software to rush development rather than invest in secure code~\citep{anderson2001information}. 
The economic aspects of information asymmetry and network effects explain why software developers often rush to market rather than focus on pre-release vulnerability discovery  \citep{anderson2001information}. 
However, this criticism of software developers presupposes a somewhat na{\"i}ve solution---exhaustive prerelease vulnerability discovery. 
Section~\ref{sec:halt-vul} provides some good reasons to believe there will always be another vulnerability to discover. 
If this is true, the optimal system for minimizing costs is less clear.

Vulnerabilities may be discovered prerelease or postrelease of the software. 
Prerelease discovery can be done only by the software developer. 
The prerelease vulnerability discovery costs we are primarily concerned with are delays to release and software developer time and attention.
Postrelease discovery can be conducted by anyone. 
The postrelease vulnerability discovery costs we are primarily concerned with are developing remediations and mitigations (such as software updates), deploying remediations and mitigations, and incidents caused by attackers exploiting the vulnerability. 
While developers bear all the prerelease costs, their customers bear the postrelease costs of deploying patches and any incidents \citep{householder2021lucky}.  

Whether the person doing vulnerability discovery is an attacker or defender, they have access to and use essentially the same  tools: program verification, fuzzing, and manual code review. 
Program verification is often easier with source code rather than compiled machine code, possibly giving close-source developers a small advantage. 
But in general, if there is an openly available program verification or fuzzing technique, then a software developer should use it because some attacker will likely use any readily available techniques to find flaws. 

Attackers have one additional method for creating exploits besides vulnerability discovery---reverse engineering security updates. 
Reverse engineering security updates does not discover new vulnerabilities, but it changes information flows. 
Attackers can often reverse a patch and start attacking the old version long before every deployed system has applied the patch or upgrade. 
Because of this lag between patch publication and patch application, there are expected incident costs to defenders even for post-release vulnerabilities discovered by \emph{defenders}.   

Whatever methods attackers use to discover vulnerabilities, an important question is how the defense community might incentivize them to stop.
In other words, make the value of some other behavior exceed the value of vulnerability discovery.
Vulnerability discovery methods such as program verification and fuzzing require compute time. 
If the economic return on that compute time when directed at vulnerability discovery is lower than that of mining bitcoin, then we expect attackers to stop spending their limited compute resources on vulnerability discovery.
Perhaps reverse engineering security patches would remain economically viable when searching for new vulnerabilities is not, but let us put that aside for a moment.
As fewer people search for previously unknown vulnerabilities, basic supply and demand predicts the value of finding them goes up as long as demand remains constant \citep{szidarovszky1977new}. 
This reasoning suggests that if there is a market for vulnerabilities there will always be people who can extract value by finding them.
Specifically, as the supply goes down, the price will rise to a point where it is worth some attacker's time to discover vulnerabilities.

There is some evidence that prices for vulnerabilities respond to market forces, with caveats that data on exploit prices is partial.
For example, observed asking price for vulnerabilities used by ransomware went down over a period of a year when there was a large increase in supply of relevant vulnerabilities in 2013~ \citep{lee2017spillover}.
Analysis of advertised buying price by vulnerability brokers from 2016 to 2022 suggest prices are most influenced by the functionality afforded by the exploit; however, the cause and effect relationship between prices and vulnerability properties remains an area for future work~\citep{dellago2022characterising}.
The computing theory described in this article may supply a hypothesis for partially explaining vulnerability market prices to examine in future work.  

The question of whether there are always more undiscovered vulnerabilities can be viewed as a market dynamics question. 
Economics often studies elasticity in regard to demand.
Inelastic demand refers to products whose consumption does not vary (much) as price increases, such as fuel. 
If vulnerabilities are ``dense,'' then this might be modeled as inelastic \emph{supply}, where there is limited variation in units produced regardless of price \citep{layton2016economics}. 
Land is a good example of a good with inelastic supply, as generally regardless of the price for land more cannot be produced. 
Vulnerabilities may have a strange place in this model because Section~\ref{sec:vul-density} suggests that \emph{fewer} vulnerabilities cannot be produced.
``Produced'' in this case means accidentally inserted during software development; insider attacks intentionally inserting malicious code are out of scope for this analysis.   
Policies that discourage discovery by attackers, such as criminalization, would not reduce supply because supply is inelastic in regards to price.
Also, intuitively, attackers are not creating the vulnerable software, simply discovering flaws in already existing software, so punishing attackers will not influence software development. 
However, such policies may reduce some specific attacker's inventory. 
While such policies may drive prices higher to compensate for the increased risk of punishment, they would not be able to quash supply of vulnerabilities available to be discovered.  

Costs borne by defenders, both system owners and suppliers, are also important.
There are several aspects to cost in vulnerability management given a known vulnerability: vulnerability risk, change risk, labor costs, and operational downtime costs \citep{cebula2010taxonomy}.
Vulnerability risk is the negative consequence of not mitigating or remediating the vulnerability.
Change risk is any unforeseen negative consequence of applying the mitigation or remediation.
Both of these are difficult to reason about because most vulnerabilities and most changes are low impact. 
Most vulnerabilities are not exploited, and most changes are managed properly via regression testing, incremental roll outs, etc. such that they do not break an organization's infrastructure. 
But some vulnerabilities let in ransomware that freezes the oil pipeline for four days, and some updates flag critical Windows DLLs as malware and render the machine inoperable. 
Labor and downtime costs are more predictable, but not negligible. 
If the cost of creating or applying a patch is greater than the cost of not creating or applying the patch, then good software management practices will not happen. 
 
An efficient software security update delivery methodology will remain important if there will always be another vulnerability. 
Efficiency is important at all levels, from speed of development through to low cost to the end user for quick deployment. 
This suggests some software designs and development methods will be better suited to deliver efficient software security updates than others.
These should be identified and preferred. 
As the SolarWinds incident demonstrated, software updates are also a risk as an attack vector. 
Due care is therefore required. 

Finally, how to prioritize use of limited resources for vulnerability management remains an open and important question. 
If there will always be more vulnerabilities, then the technical details of each specific vulnerability matter less and the context of which vulnerabilities are actively being exploited by attackers to cause material harm matters more. 
This preference is reflected in some vulnerability prioritization systems, such as the Stakeholder-Specific Vulnerability Categorization (SSVC)~\cite{spring2020ssvc}. 

\section{Conclusions}
\label{conclusion}

Vulnerabilities appear to be dense, not in the proper mathematical sense but in the sense that there will always be more vulnerabilities in a given piece of software. 
While vulnerabilities can be reduced, and best practices for secure software development need to continue, under the modern computing paradigm a pragmatically useful piece of software cannot be produced without vulnerabilities. 
This conclusion suggests three conclusions: adjust defender expectations, adjust defender practice, and consider drastic adjustments to the modern computing paradigm. 

Defenders should expect there to continually be new vulnerabilities. 
When there is a new vulnerability in an important system, the first reaction should not be surprise or denial. 
This change in expectation does not mean we stop doing all the best practices to limit vulnerabilities in software.
The efforts to reduce and limit vulnerabilities in software have made important and valuable improvements.
And these practices not only reduce vulnerabilities but also increase system reliability and fit to intended purpose.
Nonetheless, leadership at organizations should not be surprised when software contains a previously unknown vulnerability. 
This ongoing risk should be surfaced so that decisions can account for it and an organization can own the risk in its software. 

Defender practice should shift towards resiliency and efficient response. 
This advice is not new, but defenders still, to some extent, practice a castle mentality.
Of course, defenders should maintain a border and patch \acp{CVE-ID}.
However, defenders always need to segment and build resilient networks\footnote{The term ``resilient'' in this context means ``the ability of the network to provide and maintain an acceptable level of service in the face of various faults and challenges to normal operation''~\citep{sterbenz2010resilience}.} because the defender would be justified in believing the systems contain unknown vulnerabilities. 
Ongoing response to vulnerabilities actively being exploited by adversaries should become routine so that broader strategic resilience and defense can be built in.

One more drastic conclusion the community could reach is that Turing machines are actually too powerful. 
There are other abstract computation machines that are demonstrably less powerful than a Turing machine.
The \href{http://langsec.org/spw22/}{LangSec Workshop} has been dedicated to the challenge of making languages, programs, and interfaces with reduced expressiveness practical for input-handling code and input validation since 2014. 
The LangSec approach should eliminate an important class of common implementation vulnerabilities \citep{ali2021we}. 
There are various examples of creating such languages \citep{reilly2015crema}.
Developer adoption of tools produced by the LangSec academic community is not widespread; plausible explanations for this include the economic analysis discussed in Section~\ref{sec:cost}. 
Nevertheless, it is an important demonstration of a path for using sub-Turing computational models to eliminate a class of implementation vulnerabilities while still utilizing commodity hardware and operating systems.  

While sub-Turing Complete machines are less flexible, because they are less expressive programmers can prove more properties about their implementation. 
For example, researchers in a DARPA project created software-based domain-specific languages less powerful than Turing Machines in the process of creating high-assurance aviation software \citep{fisher2017hacms}.
These domain-specific language components were part of a broader formal-methods development project that led to a transition into a working Boeing prototype helicopter.
It would be an open question as to whether the hardware should be formally made sub-Turing, or whether it would be practically adequate to use sub-Turing domain-specific software languages on existing hardware. 
There is a remaining challenge that the design of the domain-specific language may have a flaw, and in time attackers may get better at recognizing those. 
However, \citep{fisher2017hacms} documents a substantial and practically important improvement in security by using sub-Turing techniques. 

Current computer chips are designed such that arbitrary computation and arbitrary transitions are possible.
This flexibility means that the chip designer does not need to know what software will be loaded and run.
An alternative would be to make computer chips more restrictive, following a computational model of a finite state machine.
The drawback is that the community would likely have to standardize what states and transitions are possible, and there will not be worldwide consensus on this.
So different chips would likely become necessary for different purposes. 
Where flexibility remains important, one could use a field programmable gate array for customization of the hardware by the end consumer. 
Such a change would have tremendous ramifications on the current design, deployment, and maintenance of information systems.

An alternative would be to exclusive run and write software in languages that are not Turing complete. 
This would leave the software exposed to vulnerabilities in the hardware itself, such as rowhammer (CVE-2015-0565).
Researchers have proved quite creative in developing and expanding on such hardware vulnerabilities \citep{spencer2021creative}, avoiding mitigations.
Whether sub-Turing Machine domain-specific languages could be designed to adequately mitigate hardware flaws is an open question.

The drastic step of reducing computing hardware to implementations of less expressive machines may be the only way to overcome the theoretical barriers to eliminating implementation vulnerabilities in Turing machines. 
As well as the formal changes, this would require a change in the way software and hardware are developed. 
Development would require more formal specifications that are amenable to model checking and program verification. 
Just switching to a sub-Turing language is not an automatic protection against implementation vulnerabilities; the switch makes the number of vulnerable states finite but not zero.
However, with a finite number of vulnerabilities, it would become possible to design program verification techniques to eliminate all of them.
Completing this and changing development practice to use them would take enormous work. 
Even if this were successful, the security community would still have work to do for prevention and remediation of design vulnerabilities. 
Fuzzing, broadly understood, will remain a useful tool for exploring design vulnerabilities. 
Whether this shift to sub-Turing hardware would go importantly beyond what is possible with sub-Turing software languages and be economically justified would be a valuable topic for future research.

\section*{Acknowledgements}
Thanks to Allen Householder, Laurie Tyzenhaus, Art Manion, the anonymous NSPW 2021 PC members, and the Computers \& Security reviewers for constructive comments on prior drafts. 

This preprint is licensed under Creative Commons Attribution license (CC BY 4.0).

The view, opinions, and/or findings contained in this material are those of the author(s) and should not be construed as an official Government position, policy, or decision; views, opinions, and/or findings do not necessarily represent those of the Department of Homeland Security.

References herein to any specific commercial product, process, or service by trade name, trade mark, manufacturer, or otherwise, does not necessarily constitute or imply its endorsement, recommendation, or favoring.

\bibliographystyle{acm}

\begin{acronym}[CERT/CC]
    	\acro{CERT/CC}{CERT{\small{}{\textregistered}} Coordination Center\acroextra{ operated by Carnegie Mellon University}}
    	\acro{CPE}{Common Platform Enumeration\acroextra{ (by \acs{NIST})}}
    	\acro{CSIR}{Computer Security Incident Response}
    	\acro{CSIRT}{Computer Security Incident Response Team}
	\acro{CVD}{Coordinated Vulnerability Disclosure}
    	\acro{CVE}{Common Vulnerabilities and Exposures\acroextra{ (by \acs{MITRE})}}
    	\acro{CVE-ID}{\ac*{CVE} identifier}
    	\acro{CVRF}{Common Vulnerability Reporting Framework}
    	\acro{CVSS}{Common Vulnerability Scoring System\acroextra{, maintained by \acs{FIRST}}}
    	\acro{CWE}{Common Weakness Enumeration\acroextra{ (by \acs{MITRE})}}
    	\acro{DARPA}{Defense Advanced Research Projects Agency}
    	\acro{DHS}{\acs*{US} Department of Homeland Security}
    	\acro{DNS}{Domain Name System}
    	\acro{DoJ}{\acs*{US} Department of Justice}
	\acro{ELF}{Executable Linkable Format}
    	\acro{HTTP}{Hypertext Transfer Protocol\acroextra{, a standard by \acs{W3C}}}
    	\acro{IEEE}{Institute of Electrical and Electronic Engineers}
    	\acro{IETF}{Internet Engineering Task Force}
    	\acro{IDS}{intrusion detection system}
    	\acro{MITRE}{the Mitre Corporation}
	\acro{ML}{machine learning}
    	\acro{NCCIC}{\acroextra{\acs*{US} }National Cybersecurity and Communications Integration Center}
    	\acro{NIST}{National Institute of Standards and Technology\acroextra{, part of the \acs*{US} Department of Commerce}}
	\acro{NVD}{National Vulnerability Database}
    	\acro{RAM}{Random Access Memory}
    	\acro{RFC}{Request for Comments\acroextra{, standardization and informational documents published by the \ac{IETF}}}
    	\acro{SANS}[SANS Institute]{Sysadmin, Audit, Network, and Security Institute}
	\acro{TLS}{Transport Layer Security}
    	\acro{UN}{United Nations}
	\acro{US}{United States of America}
\end{acronym}

\end{document}